\documentclass[aps,prx,amsmath,amssymb,floatfix,twocolumn]{revtex4-1}
\usepackage{parskip} 
\usepackage{tikz} 
\usepackage{graphicx}
\usepackage{subfigure}
\usepackage{subcaption}
\usepackage{amsmath}
\usepackage{bbold}
\usepackage{slashed}
\usepackage{amssymb}
\usepackage{braket}
\usepackage{array}
\usepackage[colorlinks]{hyperref}
\usepackage{subcaption}
\usepackage{float}
\usepackage[margin=1in]{geometry}

\usepackage{natbib}

\begin{document}
\preprint{APS/123-QED}

\title{Screening the organic materials database for superconducting metal-organic frameworks}
\author{Alexander C. Tyner$^{1,2}$}
\author{Alexander V. Balatsky$^{1,2}$}

\affiliation{$^{1}$ Nordita, KTH Royal Institute of Technology and Stockholm University 106 91 Stockholm, Sweden}
\affiliation{$^{2}$ Department of Physics, University of Connecticut, Storrs, Connecticut 06269, USA}

\date{\today}

\begin{abstract} 
The increasing financial and environmental cost of many inorganic materials has motivated study into organic  and "green" alternatives. However, most organic compounds contain a large number of atoms in the primitive unit cell, posing a significant barrier to high-throughput screening for functional properties. In this work, we attempt to overcome this challenge and identify superconducting candidates among the metal-organic-frameworks in the organic materials database using a recently proposed proxy for the electron-phonon coupling. We then isolate the most promising candidate for in-depth analysis, C$_{9}$H$_{8}$Mn$_{2}$O$_{11}$, providing evidence for superconductivity below $100$mK. 
\end{abstract}

\maketitle
\par 
\section{Introduction}
Metal-organic frameworks (MOFS) consist of clustered metallic atoms connected to a larger organic framework, often referred to as organic ligands\cite{kitagawa2014metal,furukawa2013chemistry,czaja2009industrial,zhou2012introduction,schneemann2014flexible,kuppler2009potential,mueller2006metal,cui2016metal,james2003metal,kreno2012metal,lee2009metal,long2009pervasive,wang2009postsynthetic,kurmoo2009magnetic,cohen2012postsynthetic,wang2017metal,corma2010engineering,dang2017nanomaterials,jiao2018metal,yuan2018stable}. Importantly, it has been shown that MOFS can support the desirable electronic properties of many inorganic systems but with the advantage that, as organic systems, they can often be synthesized more rapidly and at lower cost. This situation has motivated expanded experimental and theoretical studies of MOFS. 
\par 
While such studies have made significant progress, high-throughput screenings for desirable properties such as magnetism and superconductivity have fallen behind those performed for inorganic systems\cite{saal2013materials,jain2013commentary}. This is due in large part to the relatively large number of atoms, often between 20 and 200, contained within the unit cell of a typical MOF. This large number of atoms translates to a soaring computational expense in \emph{ab initio} simulations\cite{rosen2022high}. 
\par 
The computational demands of these systems are perhaps greatest when studying the phonon spectra, in particular the electron-phonon coupling\cite{RevModPhys.89.015003}. The evaluation of electron-phonon coupling strengths poses a tremendous computational challenge in simple inorganic systems, let alone large organic molecules. Nevertheless, as one of the most common mechanisms for superconductivity its evaluation is of vital importance to determining optimal superconducting candidates\cite{RevModPhys.89.015003}. 
\par 
Modern machine learning techniques appear perfectly positioned to overcome the enormous computational expense associated with studying organic compounds\cite{cgcnn,cgcnn2,schmidt2019recent}. However such methods require large datasets to be trained for making accurate predictions which do not currently exist for MOFS. 

\begin{figure}
    \centering
    \includegraphics[width=8cm]{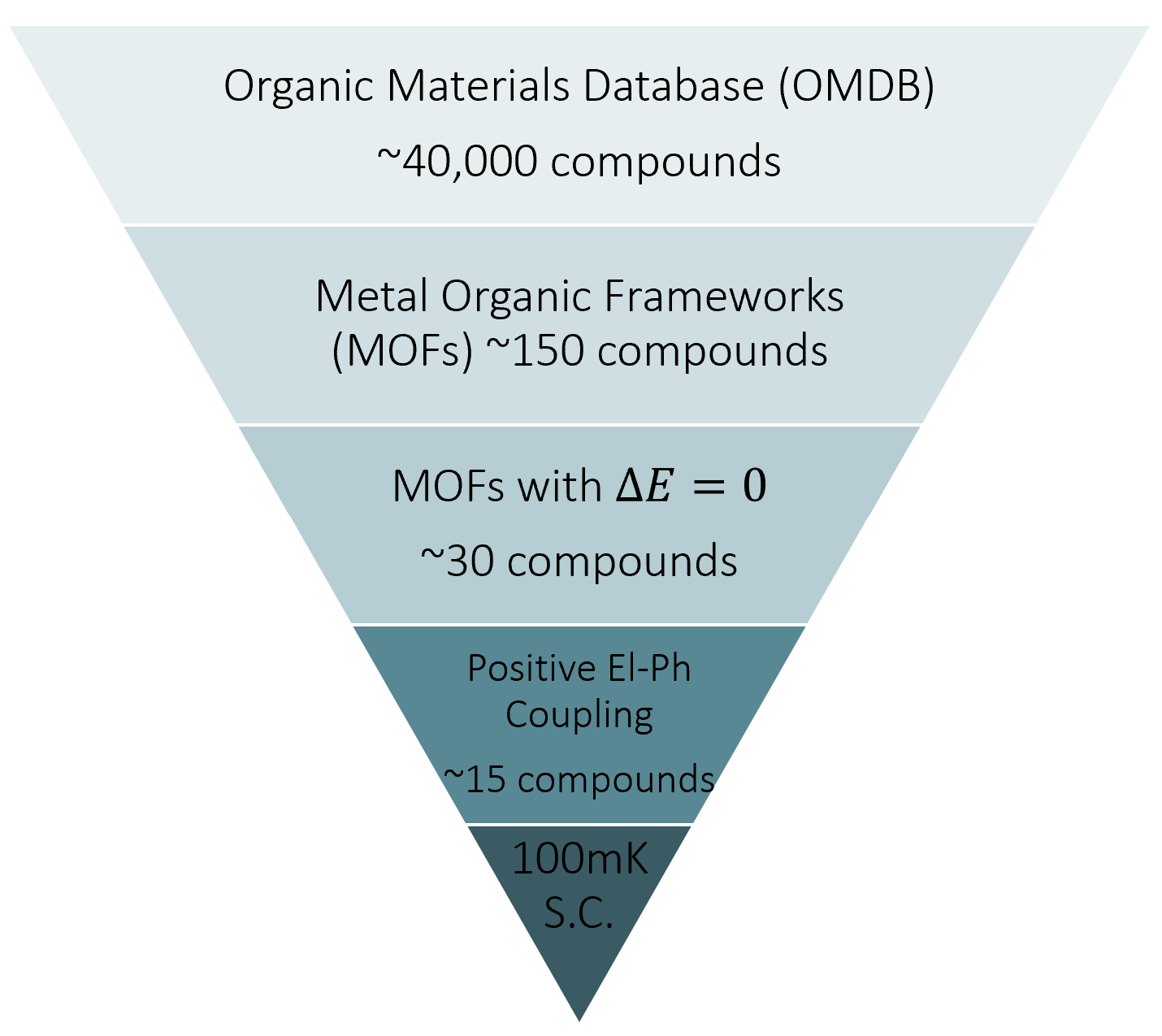}
    \caption{Methodology for isolating superconducting metal organic frameworks within the organic materials database presented in this work.}
    \label{fig:Flow}
\end{figure}
\par 
In this work, we attempt to make progress towards the screening of MOFS for superconductivity by quantifying the electron-phonon coupling, developing a strategy to make such a search feasible for the first time. In order to carry out this screening while limiting the computational expense, we utilize a recently proposed proxy for the electron-phonon coupling strength\cite{elphproxy}, $\lambda$. The fundamental idea behind this approach is that, 
\begin{equation}
    \lambda\approx f \lambda_{\Gamma},
\end{equation}
where $\lambda_{\Gamma}$ is the electron-phonon coupling strength at the $\Gamma$ location and it is shown that $f$ can reasonably be considered constant among a given material family. In this way promising superconducting candidates can be isolated by evaluating the electron-phonon coupling strength at the $\Gamma$ location alone. 
\par 
It is important to emphasize that, despite use of this proxy a large computational expense remains present due to the fact that the MOFS analyzed in this work have between 60 and 100 atoms in the primitive unit cell. To obtain suitable accuracy in \emph{ab initio} computation of the electron-phonon coupling strength for each phonon mode thus requires $\approx \times 10^{4}$ CPU hours per compound. 

\par 
In section II we outline the process by which we have filtered candidate MOFS from the organic materials database (OMDB)\cite{borysov2017organic,omdb5,omdb4,omdb2,omdb1}. The OMDB is among the largest existing databases of organic compounds, containing approximately $40000$ compounds as well as the corresponding band-structure, magnetization, and other relevant properties as computed via high-throughput density functional theory. We further describe the computational strategy employed to identify isolate the compound most likely to support superconductivity.
\par 
In section III, direct computation of $\lambda$ and the superconducting critical temperature, $T_{c}$, for this optimal compound, C$_{9}$H$_{8}$Mn$_{2}$O$_{11}$ is detailed. 
\begin{figure}
    \centering
    \includegraphics[width=8cm]{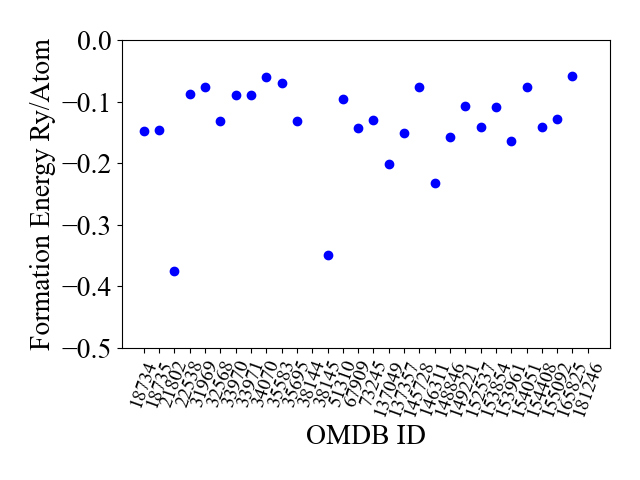}
    \caption{Formation energy of initial candidate MOFs as labelled by the ID of the compound in the organic materials database.}
    \label{fig:FormEnergy}
\end{figure}
\section{Candidate selection and evaluation}
We select MOFS from the organic materials database (OMDB). This database is extremely useful as it allows for searches to be performed based on keywords and band gap. These functionalities allow for immediate isolation of $~150$ MOFS, of which we select 30 with a vanishing band gap at the Fermi energy and less than 120 atoms in the primitive unit cell. The band-gap criteria is set as the goal of the project is to identify superconducting candidates which require a Fermi surface. The criteria for number of atoms is imposed due to considerations surrounding the computational expense of relaxation and computing phonon modes. 
\begin{figure}
    \centering
    \includegraphics[width=8cm]{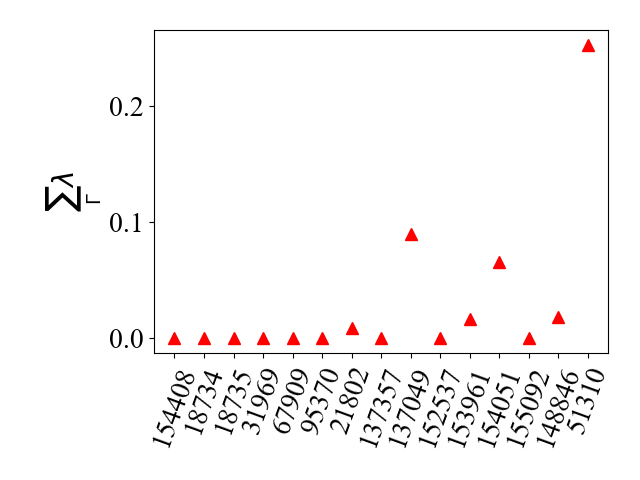}
    \caption{Electron-phonon coupling at the $\Gamma$ location for 15 candidate MOFs in the Organic Materials database as labelled by the compound ID number in the database. Compounds with non-zero values are potential superconducting candidates.}
    \label{fig:ElPh}
\end{figure}

\begin{figure*}
    \centering
    \includegraphics[width=16cm]{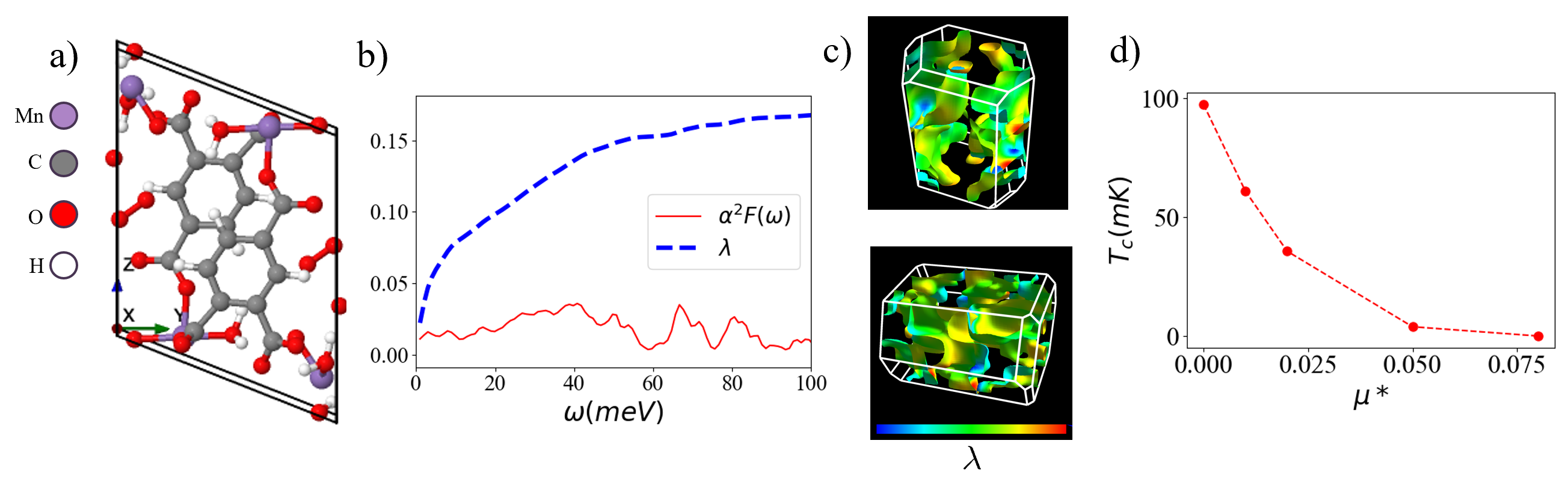}
    \caption{\textbf{Details of C$_{9}$H$_{8}$Mn$_{2}$O$_{11}$:} (a) Relaxed crystal structure of C$_{9}$H$_{8}$Mn$_{2}$O$_{11}$ for a single primitive unit cell. (b) Spectral function and electron-phonon coupling, $\lambda=\int \alpha^{2} F(\omega)/(2\omega)d \omega$. (c) Electron-phonon coupling on the Fermi surface of C$_{9}$H$_{8}$Mn$_{2}$O$_{11}$. The low-symmetry of the crystal leads to a complex Fermi surface structure. (d) Superconducting critical temperature as a function of the effective coulomb repulsion strength, $\mu * $. }
    \label{fig:51310}
\end{figure*}
\par 
Once the initial 30 candidates have been selected, the atomic positions are relaxed. In this work, all first principles calculations based on density-functional theory (DFT) are carried out using the Quantum Espresso software package \cite{QE-2009,QE-2017,QE-2020}. Exchange-correlation potentials use the Perdew-Burke-Ernzerhof (PBE) parameterization of the generalized gradient approximation (GGA) \cite{Perdew1996}. In each case the initial atomic positions and lattice parameters are taken from the OMDB. The atomic positions are relaxed until the maximum force on each atom is less than $1\times 10^{-4} \text{Ry/Bohr}$. A 6 x 6 x 6 Monkhorst-Pack grid of k-points is utilized as well as a plane wave cutoff of 520 eV.  We have neglected inclusion of a Hubbard U in these computations, which can impact the computed values and should be a subject of future work. 
\par
The formation energy of each compound, computed as $E_{compound}- \sum_{Atoms}E_{A}$, is plotted in Fig. \eqref{fig:FormEnergy}. After relaxation we proceed with computation of the electron-phonon coupling at the $\Gamma$ location. This computation is performed utilizing density functional perturbation theory (DFPT) as offered through the Quantum Espresso package. It is important to utilize DFPT in computing the phonon modes as the size of the primitive unit cell makes computation of the phonon modes via packages such as Phonopy\cite{togo2023first} which utilize supercells, computationally prohibitive.  
\par 
From the initial 30 compounds, the resulting phonon modes at the $\Gamma$ location are converged for 15 candidates. This result does not necessarily indicate that the remaining compounds are unstable, rather it is likely that the relaxation criteria for these compounds must be made more strict to achieve convergence. A common feature among those systems which do not show convergence is a large number of hydrogen atoms. Hydrogen atoms can impede the relaxation process due to their small atomic weight as incremental changes in their positions can push the system into local energetic minima.  
\par
Due to the computational expense associated with a subsequent round of relaxation and computation of phonon modes, we discard these compounds. The resulting electron-phonon coupling strength at the $\Gamma$ location for the 15 converged compounds is shown in Fig. \eqref{fig:ElPh}. 
\par
We observe that the majority of screened compounds yield a near-vanishing electron-phonon coupling strength at the $\Gamma$ location, indicating that they not likely superconductor candidates. Three compounds appear to support $\lambda_{\Gamma}>0.05$, with one compound, C$_{9}$H$_{8}$Mn$_{2}$O$_{11}$ yielding $\lambda_{\Gamma}\approx 0.28$. This value is in-line with the electron-phonon coupling strength of other three-dimensional organic superconductors\cite{zhang2017theoretical} and is thus a promising candidate warranting further investigation in the following section. 

\section{Analysis of C$_{9}$H$_{8}$Mn$_{2}$O$_{11}$}
\par
Due to the large electron-phonon coupling strength at the $\Gamma$ location relative to other candidate MOFs, we isolate C$_{9}$H$_{8}$Mn$_{2}$O$_{11}$ for further investigation. This compound belongs to symmetry class P$_{1}$ and has been experimentally synthesized with the details given in Ref. \cite{mahata2008role}. As stated previously, we utilize the atomic positions as listed in the OMDB and perform a subsequent relaxation. In order to directly investigate whether this compound is a superconducting candidate we compute the dynamical matrices within density-functional perturbation theory on an irreducible $2 \times 2 \times 2$ grid off $\mathbf{q}$-points in the Brillouin zone using DFPT as implemented in Quantum Espresso.
\par
To subsequently extract the superconducting properties of the system, we implement Migdal-Eliashberg theory via the Electron-Phonon Wannier (EPW) code\cite{lee2023electron,PhysRevB.76.165108,PhysRevB.87.024505}. This is an advantageous approach as the EPW code uses Wannier interpolation to access significantly denser $\mathbf{k}$ and $\mathbf{q}$ point grids in computation of the electron-phonon matrix elements. The Wannier interpolation is performed using a $8 \times 8 \times 8$ grid of $\mathbf{k}$ points. To achieve optimal selection of atomic orbitals, we implement the SCDM method\cite{vitale2020automated,Pizzi2020}. We subsequently utilize $10 \times 10 \times 10$ $\mathbf{k}$ and $\mathbf{q}$ point grids for computation of the electron-phonon matrix elements.

\par 
The resulting Eliashberg spectral function is shown in Fig. \eqref{fig:51310}(b) as well as the resulting electron-phonon coupling strength, 
\begin{equation}
    \lambda(\omega)=\int \alpha^2 F(\omega)/(2\omega)d\omega.
\end{equation}
We find that $\lambda$ saturates to maximum value of $\lambda\approx 0.19$. We further demonstrate a plot of $\lambda$ on the Fermi surface in Fig. \eqref{fig:51310}(c). This plot underscores the low-symmetry which is common among organic systems. We further estimate the superconducting critical temperature ($T_{c}$) by self-consistently solving the isotropic Migdal-Eliashberg equations for multiple effective coulomb repulsion values with the results shown in Fig. \eqref{fig:51310}(d).  
\par 
The estimated values of $T_{c}$ imply that, while posing a non-trivial challenge, the superconducting state can be accessed experimentally. It should be emphasized that we have neglected to include both spin-orbit coupling and a Hubbard U in this computation. Due to the presence of Mn both should effect the results of the realistic system and  magnetic pairing mechanism cannot be ruled out\cite{monthoux2007superconductivity}. We further point out that recent studies have identified significant changes in the critical temperature when approaching the two-dimensional limit in MOFs\cite{zhang2017theoretical}. It is possible that such a dependence can be identified in C$_{9}$H$_{8}$Mn$_{2}$O$_{11}$. Investigation of these effects to future works.

\section{Summary and outlook}
Our work validates the utility of the method proposed in Ref. \cite{elphproxy} in organic compounds and demonstrates an effective route to screening organic compounds for superconducting candidates. With the necessary computing resources this approach can be scaled up to screen all known MOFs and identify optimal candidates. This is an important step forward given the current environmental and financial cost of known superconductors composed of rare earth metals. Our work further puts forward a novel, stable organic superconductor, C$_{9}$H$_{8}$Mn$_{2}$O$_{11}$ motivating future experimental studies.

\acknowledgements{}
We thank Sin\'ead M. Griffin for stimulating discussions and drawing our attention to Ref. \cite{elphproxy}. We acknowledge support from the European Research Council under the European Union Seventh Framework ERS-2018-SYG 810451 HERO and the University of Connecticut. Nordita is supported in part by NordForsk. The computations were enabled by resources provided by the National Academic Infrastructure for Supercomputing in Sweden (NAISS), partially funded by the Swedish Research Council through grant agreement no. 2022-06725.

\bibliography{ref.bib}
\end{document}